\documentclass{PoS}
\usepackage{cite}
\usepackage{epsfig}

\title{
\vspace*{-2.3cm}
\begin{minipage}{\textwidth}
{\normalfont\small LTH 938, DESY 12-029, LPN12-037, SFB/CPP-12-12
\hspace{\fill} February 2012}\\
\end{minipage}\\[35pt]
        \boldmath Generalized threshold resummation for semi-inclusive
        e$^{+}$e$^{-}$ annihilation}
 
\ShortTitle{Generalized threshold resummation for semi-inclusive $e^+e^-$ 
annihilation}

\author{\speaker{N.A. Lo Presti}\\
        Department of Mathematical Sciences, University of Liverpool \\
       Liverpool L69 3BX, United Kingdom\\
        E-mail: \email{lopresti@liverpool.ac.uk}}
        
\author{A.A. Almasy\\
        Deutsches Elektronensynchrotron DESY \\
    Platanenallee 6, D--15738 Zeuthen, Germany\\
        E-mail: \email{andrea.almasy@desy.de}}

\author{A. Vogt\\
        Department of Mathematical Sciences, University of Liverpool \\
        Liverpool L69 3BX, United Kingdom\\
        E-mail: \email{Andreas.Vogt@liverpool.ac.uk}}        
        
\abstract{
\vspace*{5mm}
Recently methods have been developed to extend the resummation of large-$x$
double logarithms in inclusive deep-inelastic scattering (DIS) to terms not
addressed by the soft-gluon exponentiation. Here we briefly outline our 
approach based on fixed-order results, the general large-$x$ structure in 
dimensional regularization and the all-order factorization of mass 
singularities, which is directly applicable also to semi-inclusive $e^+e^-$
annihilation (SIA). We then present some main results for the corresponding
timelike splitting functions and transverse and longitudinal fragmentation 
functions. The close relation between DIS and SIA facilitates the 
determination of additional third-order results for the latter function which 
is fully known only at the next-to-leading order. Therefore all above 
quantities can be resummed at next-to-next-to-leading logarithmic accuracy.}

\FullConference{ 10th International Symposium on Radiative Corrections
(Applications of Quantum Field Theory to Phenomenology) - Radcor2011\\
September 26-30, 2011, 
Mamallapuram, India}

\newcommand{\lsim}{\raisebox{-0.07cm}{$\:\:\stackrel{<}{{\scriptstyle
 \sim}}\:\: $} }

\newcommand{\hspn}{{\hspace{-4mm}}}

\newcommand{\hspp}{{\hspace{5mm}}}
\newcommand{\beq}{\begin{equation}}
\newcommand{\eeq}{\end{equation}}
\newcommand{\bea}{\begin{eqnarray}}
\newcommand{\eea}{\end{eqnarray}}
\newcommand{\nn}{\nonumber}

\newcommand{\MSb}{$\overline{\mbox{MS}}$}
\newcommand{\as}{\alpha_{\rm s}}
\newcommand{\ar}{a_{\rm s}}
\newcommand{\art}{\tilde{a}_{\rm s}}

\newcommand{\ep}{\epsilon}

\newcommand{\Ntil}{\widetilde{\!N}}
\newcommand{\GE}{\gamma_{\rm e}}

\def\frct#1#2{\mbox{\large{$\frac{#1}{#2}\:\!$}}}
\newcommand{\eqM}{\raisebox{-0.07cm}{$\stackrel{\wedge}{=} $} }

\begin{document}

\def\Ftwo{{F_{\, 2}}}
\def\FL{{F_{\:\! L}}}
\def\F3{{F_{\:\! 3}}}
\def\Qs{{Q^{\, 2}}}
\def\GeV2{{\mbox{GeV}^{\:\!2}}}
\def\x1{{(1 \! - \! x)}}
\def\z#1{{\zeta_{\:\! #1}}}
\def\ca{{C_A}}
\def\cas{{C^{\: 2}_A}}
\def\cat{{C^{\: 3}_A}}
\def\caaf{{C^{\: 4}_A}}
\def\canm{{C^{\: n-1}_A}}
\def\canmm{{C^{\: n-2}_A}}
\def\canmmm{{C^{\: n-3}_A}}
\def\canlm{{C^{\: n-\ell-1}_A}}
\def\canlmm{{C^{\: n-\ell-2}_A}}
\def\canlmmm{{C^{\: n-\ell-3}_A}}
\def\cf{{C_F}}
\def\cfs{{C^{\: 2}_F}}
\def\cft{{C^{\: 3}_F}}
\def\cff{{C^{\: 4}_F}}
\def\cffi{{C^{\: 5}_F}}
\def\cfl{{C^{\: \ell}_F}}
\def\cfn{{C^{\: n}_F}}
\def\cfnm{{C^{\: n-1}_F}}
\def\cfnmm{{C^{\: n-2}_F}}
\def\nf{{n^{}_{\! f}}}
\def\nfs{{n^{\,2}_{\! f}}}
\def\nft{{n^{\,3}_{\! f}}}
\def\caf{{C_{AF}}}
\def\cafs{{C_{AF}^{\: 2}}}
\def\caft{{C_{AF}^{\: 3}}}
\def\caff{{C_{AF}^{\: 4}}}
\def\cafn{{C_{AF}^{\: n}}}
\def\cafnm{{C_{AF}^{\: n-1}}}
\def\cafnmm{{C_{AF}^{\: n-2}}}
\def\dabc2{{d^{\:\!abc}d_{abc}}}
\def\dabcnc{{{d^{\:\!abc}d_{abc}}\over{n_c}}}
\def\dabcna{{{d^{\:\!abc}d_{abc}}\over{n_a}}}
\def\fl11{fl_{11}}
\def\flg11{fl^g_{11}}
\def\fl02{fl_{02}}
\def\b#1{{{\beta}_{#1}}}
\def\bb#1#2{{{\beta}_{#1}^{\,#2}}}

\def\B#1{{{\cal B}_{\:\!#1}}}

\section{Introduction}

\vspace*{-3mm}
\noindent
In the past years quite a few studies have addressed the threshold behaviour of
higher-order splitting functions and hard-process coefficient functions in 
perturbative QCD beyond the quantities or contributions covered by the standard
soft-gluon exponentiation
\cite{Laenen:2008ux,Laenen:2008gt,Gardi:2010rn,Laenen:2010uz,Grunberg:2009yi,%
Grunberg:2009vs,Grunberg:2011gx,MV3,MV5,SMVV1,SMVVpr,AV2010,ASV1}. 
Except for the gluon contributions to the longitudinal structure and 
fragmentation functions $F_L$ which include an additional factor $\x1$, the 
dominant terms are of the divergent but integrable `double-logarithmic' form 
\beq
\label{doublogs}
  \as^{\,n} \ln^{\:\!2n-n_0^{}-\ell}\!\x1 \; .
\eeq 
Here $n_0^{}$ depends on the quantity under consideration, with $n_0^{} = 2$ 
for the `off-diagonal' splitting functions $P_{\, ij}(x)$ with $i\neq j$
-- recall that the diagonal splitting functions $P_{\, ii}(x)$ do not exhibit 
any $n$-dependent large-$x$ enhancement in the \MSb\ scheme adopted in the 
present contribution \cite{Korchemsky:1989si,DMS05} --
and $n_0^{} = 1$ for the structure functions in deep-inelastic scattering (DIS) 
and fragmentation functions in semi-inclusive $e^+e^-$ annihilation (SIA)
with the exception of the functions $F_L$ where $n_0^{} = 2\,$. 
$\ell = 0,\: 1,\: 2,\:\ldots$ in Eq.~(\ref{doublogs}) represent the 
leading logarithmic (LL), next-to-leading logarithmic (NLL),
next-to next-to-leading logarithmic (NNLL = N$^2$LL), \dots\ large-$x$ 
contributions.

Using the third-order results of Refs.~\cite{MVV6,MVV10},
the LL, NLL and NNLL coefficients of the (non-singlet) quark coefficient 
functions for the structure functions $F_2$, $F_3$ and $F_L$ in DIS and their 
counterparts $F_T$, $F_A$ and $F_L$ in SIA have been obtained in Refs.~\cite
{MV3,MV5}, together with the corresponding LL and NLL results for the 
Drell-Yan process, from the single-logarithmic behaviour of the physical 
evolution kernels of these observables.
The corresponding approach is not sufficient for all-order predictions in  
flavour-singlet cases, but was used to obtain the NNLL approximations of the 
fourth-order contributions to the off-diagonal spacelike splitting functions 
$P_{ij}^{\,S}$ for the parton distribution and the longitudinal coefficient 
function $C_{L,g}$ in DIS \cite{SMVV1,SMVVpr}.

Those DIS results have been confirmed and extended by studying the 
$D$-dependence of the unfactorized structure functions in dimensional 
regularization \cite{AV2010,ASV1}. This approach facilitates the determination 
of the one previously missing parameter in the N$^3$LL non-singlet coefficients
for $F_2$ and $F_3$ -- a result that has been obtained independently in Ref.~%
\cite{Grunberg:2009vs} -- and the extension of the NNLL all-order resummation 
to the off-diagonal splitting functions and gluon coefficient functions.
\linebreak
In the present contribution we briefly report on the generalization of those 
results to the off-diagonal timelike splitting functions $P_{\, ij}^{\,T}$ for 
parton fragmentation and the singlet coefficient functions for the
fragmentation functions in SIA; a detailed account of our results will be
published elsewhere \cite{ALPVprp}.

For this purpose we consider the transverse and longitudinal gauge-boson
exchange fragmentation functions $F_T^T$ and $F_L^T$ as defined in 
Ref.~\cite{PDG2010} together with the corresponding Higgs-decay quantity in 
the heavy top-quark limit $F_\phi^T$. 
Schematically our notation for the corresponding timelike coefficient functions 
$C_{a,i}^{\,T}$ and the evolution of the fragmentation distributions $D_i$ at 
the physical scale $\Qs$ with $\ar = \as(\Qs)/(4\pi)$ reads (note the 
transposition of the splitting-function matrix)
\beq
\label{CPtime}
  F_a^{\,T}(\Qs) \:=\: \,C_{a,i}^{\,T}(\ar) \otimes D_i(\Qs)
  \;\; , \quad
  \frac{d}{d\ln \Qs}\: D_j \;=\; P^{\,T}_{ij}(\ar) \otimes D_i 
\eeq
where $\otimes$ represent the Mellin convolution and the appropriate
summations over $i$ are understood. 
 
\section{Outline of the resummation}

\vspace*{-3mm}
\noindent
The primary objects of the present resummation are the unfactorized SIA
fragmentation functions
\beq
\label{FThatFact}
  \widehat{F}_{a,k}^{\,T} \;=\; \widetilde{C}_{a,\,i}^{\,T} \otimes
  Z_{ik}^{T} \; 
  \quad \mbox{for} \quad
  a,k \,=\, T,g\,,\:\: \phi,q\,,\:\: L,q\;\; \mbox{and} \;\; L,g
\eeq
in $D = 4 - 2\:\!\ep$ dimensions. The functions $\widetilde{C}_{a,\,i}^{\,T}$ 
are given by Taylor series in $\ep$, with the $\ep^k$ terms including $k$ more 
powers in $\ln\x1$ than the 4-dimensional coefficient functions. The timelike 
transition matrix $Z^T$ consist of only negative powers of $\ep$ and 
can be written in terms of
\beq
\label{Pexp}
  P^{\,T} \;=\; \ar\, P_0^{} \:+\: \ar^{\,2}\, P_1^{} 
  \:+\: \ar^{\,3}\, P_2^{} \:+\: \dots
\eeq
and the corresponding coefficients $\beta_m$ of the beta function. This 
dependence can be summarized as
\beq
\label{ZofPn}
  \ar^{\,n}\ep^{\,-n}:\; P_{\,0\,},\: \b0 \;\; , \quad
  \ar^{\,n}\ep^{\,-n+1}:\; +\, P_{\,1},\: \b1 \;\; , \quad
  \dots \;\; , \quad
  \ar^{\,n}\ep^{\,-1}:\; P_{n-1} \; \; . 
\eeq
Hence fixed-order knowledge at N$^m$LO (i.e., of the splitting functions to 
$P_m$ and the corresponding coefficient functions) fixes the first $m\!+\!1$ 
coefficients in the $\ep$ expansion of $\widehat{F}_{a,k}^{\,T}$ at all orders 
in~$\ar$.
The large-$x$ expansions of $\widehat{F}_{a\neq L,\,k}^{\,T}$ (the 
corresponding relation for $F_L$ is slightly different) are given~by
\beq
\label{FThatLogs}
  \widehat{F}^{\,T} \big|_{\ar^{\,n}\ep^{\,-n+\ell}} \;=\;
  {\cal F}_{n,\ell}^{(0)} \ln^{\:\!n+\ell-1}\x1 \:+\:
  {\cal F}_{n,\ell}^{(1)} \ln^{\:\!n+\ell-2}\x1 \:+\: \dots \;\; .
\eeq
If the constants up to ${\cal F}_{n,\ell}^{(m)}$ are known for all $n$ and 
$\ell$, then the splitting functions and coefficient functions can be 
determined at N$^m$LL accuracy at all orders of the strong coupling.

As in DIS, the $n^{\rm th}$ order large-$x$ contributions to 
$\widehat{F}_{a\neq L,\,k}^{\,T}$ are built up from $n$ term of the form
\beq
\label{FThatDec}
  \big( A_{n,k} \,\ep^{\,-2n+1} \,+\, B_{n,k} \,\ep^{\,-2n+2} 
  \,+\, C_{n,k} \,\ep^{\,-2n+3} \,+\, \ldots \big) \x1^{-k\:\! \ep} 
  \;\; , \quad k = 1,\, \dots,\, n
\eeq
which arises from the phase-space integrations for the undetected final-state
partons and the loop integrals of the virtual corrections
\cite{MvN88,MvdMvN89,RvNnpb}. Since the terms with $\ep^{\,-2n+1}, \:\dots\,,\: 
\ep^{\,-n-1}$ have to cancel in sum (\ref{FThatFact}), there are $\,n\!-\!1$ 
relations between the LL coefficients $A_{n,k}$ which lead to the constants
${\cal F}_{n,\ell}^{(0)}$ in Eq.~(\ref{FThatLogs}), $\,n\!-\!2$ relations 
between the NLL coefficients $B_{n,k}$ etc. 
As discussed above, a N$^m$LO calculation fixes the (non-vanishing) 
coefficients of $\ep^{\,-n}, \:\dots\,,\: \ep^{\,-n+m}$ at all orders $n$,
adding $m+1$ more relations between the coefficients in Eq.~(\ref{FThatDec}).
Consequently the highest $m\!+\!1$ double logarithms, i.e., the N$^m$LL 
approximation, can be determined in this manner from the N$^m$LO results.

The coefficient functions for $F_T \equiv F_T^T$, $F_L^T$ and $F_\phi^T$ are 
known at the second order \cite{RvNplb,RvNnpb,MMoch06,AMV1}. 
The third-order timelike splitting functions have been determined, up to an 
uncertainty which is irrelevant in the present context, in 
Refs.~\cite{MMV06,MV2,AMV1}; see also Ref.~\cite{Grunberg:2011gx} for their 
large-$x$ logarithms. The all-order factorization of the quantities 
(\ref{FThatFact}) requires corresponding large-$x$ results for the 
quantities $F_{T,q}^{\,T}$ and $F_{\phi,g}^{\,T}$. These are available
from the soft-gluon exponentiation to a far higher accuracy than required here,
cf.~Ref.~\cite{MV4}. The calculations are carried out in Mellin-$N$
space with
\beq
  \ln^{\,k}\x1 \:\;\eqM\;\: (-1)^k\,N^{\,-1} \big( \ln^{\,k} \Ntil
    \:+\: \frct{1}{2}\, k (k-1) \zeta_{\,2}\, \ln^{\,k-2} \Ntil 
    \:+\: \dots \big) \;\; , \;\; \ln\, \Ntil = \ln N + \GE
\eeq
where $\GE$ is the Euler-Mascheroni constant.
The required formalism is completely analogous to that in Ref.~\cite{ASV1}.
Our symbolic manipulations have been performed using {\sc Form} and
{\sc TForm} \cite{Form3,TForm}.

\setcounter{equation}{0}
\section{Results for the timelike splitting functions}
 
\vspace*{-3mm}
\noindent
As their spacelike counterparts, the LL and NLL contributions to the
off-diagonal timelike splitting functions can be expressed in terms of 
functions $\B{n}(x)$ introduced and discussed in Refs.~\cite{AV2010,ASV1},
\beq
\label{Bkdef}
  \B{k}(x) \;=\;
  \sum_{n\,=\,0}^\infty \;\frac{B_n}{n!(n+k)!}\; x^{\,n}
 \quad \mbox{and} \quad
  \B{-k}(x) \;=\;
  \sum_{n\,=\,k}^\infty \;\frac{B_n}{n!(n-k)!}\; x^{\,n} 
\eeq
for $k = 0,\:1\:,2,\:\dots$, where $B_n$ are the Bernoulli numbers in the 
normalization of Ref.~\cite{AbrSteg}. We obtain
\bea
\label{Pqgres} 
 N\:\!P^T_{\rm qg}(N,\as) &=& 2\:\!\ar\,\nf\,\B0(-\art) 
 \nn \\[-1mm] & & \mbox{\hspn}
 +\,\ar^{\,2} \ln\, \Ntil\: \nf
    \Big[\left(12\:\!\cf-6\:\!\b0\right)\:\frac{1}{\art}\:\B{-1}(-\art)
  \,-\: \frac{\b0}{\art}\:\B{-2}(-\art) 
 +\, \left(6\:\!\cf-\b0\right)\:\B1(-\art)\Big]
 \nn \\ & & \mbox{\hspn}
 +\; \mbox{ NNLL contributions } \;+\; \ldots \:\: ,
\\
\label{Pgqres}
 N\:\!P^T_{\rm gq}(N,\as) &=& 2\:\!\ar\,\cf\,\B0(\art) 
 \nn \\[-1mm] & & \mbox{\hspn} 
 +\,\ar^{\,2} \ln\, \Ntil\: \cf 
    \Big[\left(12\cf-2\b0\right)\:\frac{1}{\art}\:\B{-1}(\art)
  \,+\: \frac{\b0}{\art}\:\B{-2}(\art) 
  \,+\, \left(8\ca-2\cf-\b0\right) \B1(\art) \Big]
 \nn \\ & & \mbox{\hspn}
 +\; \mbox{ NNLL contributions } \;+\; \ldots \:\: .
\eea
Here $C_A$ and $C_F$ are the standard $SU(n_c$) colour factors with $C_A=n_c=3$ 
and $C_F=4/3$ in QCD, $\nf$ represents the number of effectively massless quark 
flavours, $\beta_0 = 11/3\;C_A - 2/3\;\nf$ is the first coefficient of the
(four-dimensional) beta function, and we have used the shorthand
\beq
\label{atilde}
  \art \;\equiv\; 4\:\!\ar\,\caf\,\ln^2 \Ntil 
  \quad \mbox{with}\quad \caf \equiv C_A-C_F
\eeq
reflecting the vanishing of the double logarithms for the `supersymmetric' 
case $C_A = C_F$. 

\begin{figure}[p]
\centerline{\epsfig{file=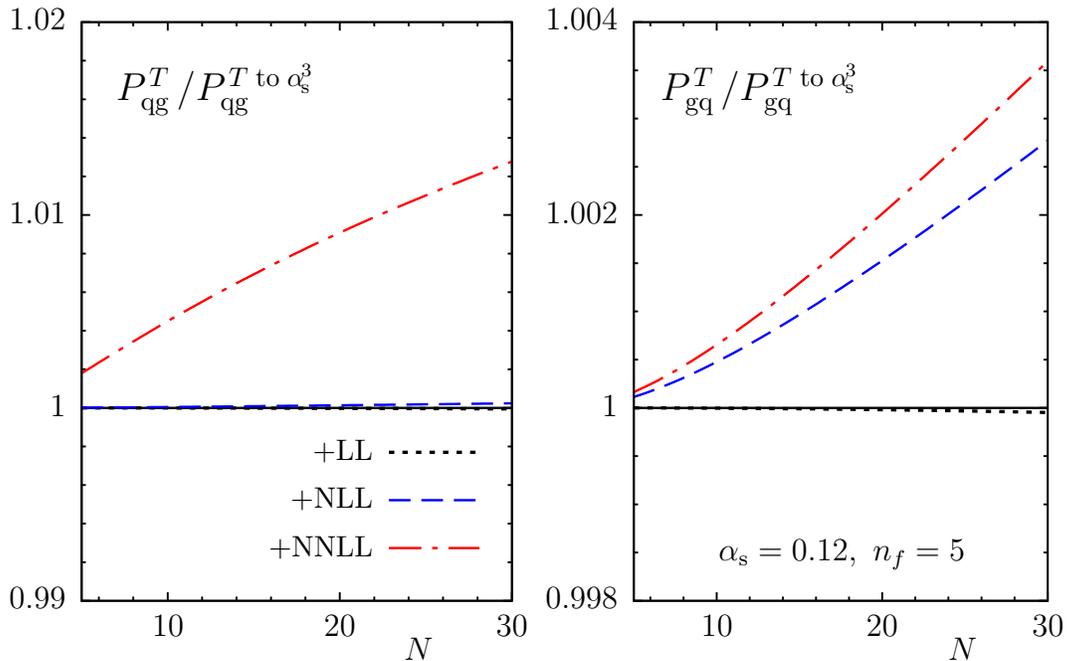,width=14cm,angle=0}}
\vspace{-1mm}
\caption{The relative leading-logarithmic (LL), next-to-leading logarithmic
(NLL) and next-to-next-to-leading logarithmic (NNLL) higher-order large-$x$ 
corrections to the third-order off-diagonal timelike splitting functions 
$P_{ij}^{\,T}$ in Mellin-$N$ space at a typical high-scale reference point.
\label{Fig1}
}
\end{figure}

The first and second lines of Eqs.~(\ref{Pqgres}) and (\ref{Pgqres}) are the 
LL and NLL results, respectively. They differ from their spacelike 
counterparts, $\nf/C_F\: P^{\,S}_{\rm gq}$ and $C_F/\nf \: P^{\,S}_{\rm qg}$ 
only by coefficients of $\B1$ proportional to $C_A-C_F$. 
These differences are due to different $(1\!-\!\ep)^{-1}$ prefactors of the 
spacelike and timelike unfactorized structure functions, e.g., the absence of 
the DIS gluon spin-averaging in the corresponding SIA quantity, which are 
otherwise related by a (at this level) simple analytic continuation. 
As in the spacelike case, we have no closed all-order expression for the NNLL 
contributions to Eqs.~(\ref{Pqgres}) and (\ref{Pgqres}) which therefore will 
be presented via tables to a sufficiently  high order in $\as$ in 
Ref.~\cite{ALPVprp}. After transformation back to $x$-space the fourth-order 
results read 
\bea
\label{Pgq3DL0}
  P_{\rm qg}^{\,T}(x)\Big|_{\ar^{\,4}} & = &
    \mbox{\hspn}
     \;\;\;\; \ln^{\,5}\! \x1\, \* \cafs \* \nf \Big[\,
            \frct{22}{27}\: \* \caf 
          \,- \, \frct{14}{27}\: \* \cf 
          \,- \, \frct{4}{27}\: \* \nf
          \Big]
\nn \\ & & \mbox{\hspn}
     + \ln^{\,4}\! \x1\, \caf \* \nf \* \Big[\,
            \Big( \, \frct{1432}{81}
            \,+\, \frct{64}{9}\, \* \z2\! \Big)\, \* \cafs 
          \,+\, \Big( \, \frct{1471}{54}
          \,-\, 8 \* \z2\! \Big)\, \* \caf \* \cf 
\\ & & \mbox{}
          \,-\, \frct{16}{3}\: \* \caf \* \nf
          \,-\, \frct{49}{81}\: \* \cfs 
          \,+\, \frct{17}{81}\: \* \cf \* \nf
          \,+\, \frct{32}{81}\: \* \nfs
          \Big]
      + {\cal O} \left( \ln^3 \! \x1 \right)
\nn \:\: , \\[2mm] 
\label{Pqg3DL0}
  P_{\rm gq}^{\,T}(x)\Big|_{\ar^{\,4}} &\! = \!&
     \mbox{\hspn}
     \;\;\;\; \ln^{\,5}\! \x1\, \* \cafs \* \cf \* \Big [ \:
          \,-\, \frct{26}{27}\: \* \caf 
          \,-\, \frct{14}{27}\: \* \cf 
          \,-\, \frct{4}{27}\: \* \nf
          \Big]
\nn \\ & & \mbox{\hspn}
     + \ln^{\,4}\! \x1\, \caf \* \cf \* \Big [ \,
            \Big( \, \frct{469}{27}
            \,-\, \frct{128}{9}\, \* \z2\! \Big)\, \* \cafs 
          \,+\, \Big( \, \frct{5317}{162}
          \,-\, 8 \* \z2\! \Big)\, \* \caf \* \cf 
\\ & & \mbox{}
          \,-\, \frct{212}{81}\: \* \caf \* \nf
          \,-\, \frct{13}{81}\: \* \cfs 
          \,+\, \frct{17}{81}\: \* \cf \* \nf
          \,-\, \frct{4}{81}\: \* \nfs
          \Big]
      + {\cal O} \left( \ln^3 \! \x1 \right)
\nn \:\: .
\eea
%
%
The LL coefficients vanish at all even orders in $\as$ from the 
fourth due to $B_{2n+1}=0$ for $n \geq 1$. \linebreak
The above results are illustrated in Fig.~\ref{Fig1} below for $\as = 0.12$ and 
$\nf = 5$, i.e., at a scale $\Qs \simeq M_Z^{\,2}$.

\begin{figure}[p]
\vspace*{2mm}
\centerline{\epsfig{file=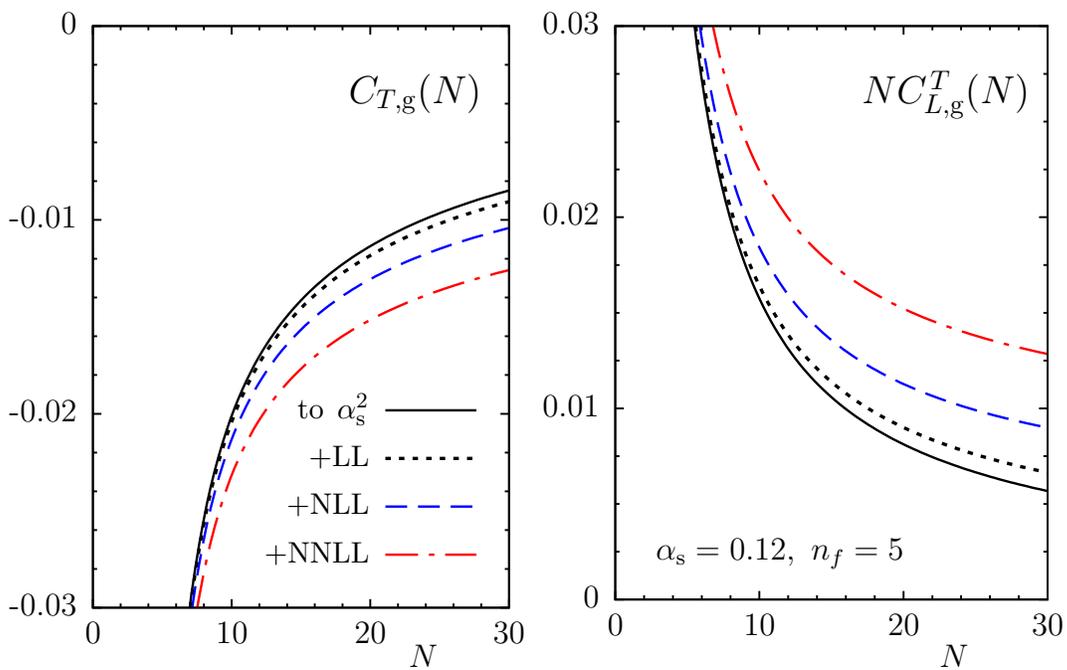,width=14cm,angle=0}}
\vspace{-1mm}
\caption{The absolute LL, NLL and NNLL higher-order 
threshold corrections to the second-order transverse and longitudinal gluon
coefficient functions $C_{T,\rm g}$ and $C_{L,\rm g}^{\:T}$ in $N$-space at a
scale $\Qs \simeq M_Z^{\:2}$.
Note that our normalization of both functions differs by a factor of 
$\frac{1}{2}$ from that of Refs.~\cite{RvNnpb,RvNplb,MMoch06}, i.e., here the 
lowest-order large-$x$ limits are 
$\,C_{T,g}(x,\ar)       \,=\, 2\,\cf \ar \, \ln \x1 \,+\, \ldots$ and 
$\,C_{L,g}^{\,T}(x,\ar) \,=\, 4\,\cf \ar \, (1-x) \,+\, \ldots\;\; $.
\label{Fig2}
 }
\vspace*{-1mm}
\end{figure}

\setcounter{equation}{0}
\section{Results for the SIA coefficient functions}

\vspace*{-3mm}
\noindent
The NNLO results facilitate the resummation of also the (more complicated) 
coefficient functions $C_{T,\rm g}$ and $C^{\,T}_{\phi,\rm q}$ at 
NNLL accuracy. Here we show, for brevity, only the NLL resummation of the 
former quantity (the NNLL contributions and the result for the latter will be 
given in Ref.~\cite{ALPVprp}),
\bea
 N\:\!C_{T,\rm g}(N,\as) &\!=\! & 
  \frac{1}{2\ln \,\Ntil^{}}\: \frac{\cf}{\ca-\cf}
  \left[\exp(2\ar\cf\ln^2\Ntil^{})\B0(\art)-\exp(2\ar\ca\ln^2\Ntil)\right]
\nn \\[0mm] & & \mbox{\hspn} -\: 
  \frac{1}{8\ln^2\Ntil}\: \frac{\cf(3\cf-\b0)}{(\ca-\cf)^2} 
  \left[\exp(2\ar\cf\ln^2\Ntil)\B0(\art)-\exp(2\ar\ca\ln^2\Ntil)\right]
\\[0mm] & & \mbox{\hspn} -\: 
  \frac{\ar}{4}\: \frac{\cf}{\ca-\cf}\: 
  \exp\,(2\ar\ca\ln^2\Ntil)\left(8\ca+4\cf-\b0\right) 
\nn \\[0mm] & & \mbox{\hspn} -
  \frac{\ar^2}{3}\,\b0\ln^2 \Ntil\,\frac{\cf}{\ca-\cf}
  \left[\ca\exp(2\ar\ca\ln^2\Ntil) - \cf\exp(2\ar\cf\ln^2\Ntil)\B0(\art)\right]
\nn \\[-1mm] & & \mbox{\hspn} -\:
  \frac{\ar}{4}\:\frac{\cf}{\ca-\cf}\: 
  \exp\,(2\ar\cf\ln^2\Ntil)\Big[-6\cf\B0(\art)-(8\ca-2\cf-\b0)\B1(\art) 
\nn \\[-1.5mm] & & \mbox{\hspp} -\:
  (12\cf-4\b0)\:\frac{1}{\art}\:\B{-1}(\art)-\:\frac{\b0}{\art}\:\B{-2}(\art)
  \Big] 
  \:+\: \mbox{ NNLL contributions } \;+\; \ldots \:\: . \nn
\label{eq:c2gNL}
\eea
Also this expression differs from its spacelike counterpart in Ref.~\cite{ASV1}
only in the coefficient of~$\B1$. The third-order contribution, now including
the NNLL term, is given by
\bea
\label{cg3DL3}
 C_{T,{\rm g}}(x)\Big|_{\ar^{\,3}} & = & 
    \mbox{\hspn}
     \;\;\;\; \ln^{\,5}\! \x1 \* \,\cf \* \Big[\,
           \frct{2}{3} \*\cas
          \,+\, \frct{10}{3} \* \cfs
          \Big]
\nn \\ & & \mbox{\hspn}
     + \ln^{\,4}\! \x1 \* \,\cf \* \Big[\,
            \frct{7}{27} \* \ca \* \nf
          \,-\, \frct{269}{54} \* \cas
          \,+\, \frct{17}{27} \* \cf \* \nf
          \,-\, \frct{338}{27} \* \cf \* \ca
          \,-\, \frct{97}{18} \* \cfs
          \Big]
\nn \\[0.5mm] & & \mbox{\hspn}
     + \ln^{\,3}\! \x1 \* \,\cf \* \Big[
          \Big(\, \frct{2990}{81}
            -   \frct{16}{9}\, \* \z2\!\Big)\,\* \cas
          \,+\,\Big(\, \frct{3652}{81}
            -   \frct{88}{9}\, \* \z2\!\Big)\,\* \cf \* \ca
\nn \\[0.5mm] & & 
          \,-\, \Big(\,\frct{41}{9}
            -   \frct{112}{9}\, \* \z2\!\Big)\,\* \cfs
          \,-\, \frct{140}{81} \* \ca \* \nf
          \,-\, \frct{436}{81} \* \cf \* \nf
          \Big]
\;\;+\;\; {\cal O} \left( \ln^2 \! \x1 \right) \:\: .
\eea

For the longitudinal fragmentation function the second-order results represent 
only the NLO contribution. The resulting NLL results for the gluon coefficient 
function read 
\bea
N^{\,2}\:\!C^T_{L,\rm g}(N,\as) & = & \:
 4\:\!\ar\,\cf\,\exp\,(2\:\!\ar\ca\ln^2 \Ntil)
 \;+\; 2\:\!\ar\cf\, N\:\! C^{\,\rm LL}_{T,\rm g}(N,\as)
\nn \\[-1mm] & & \mbox{\hspn} +\:
 8\:\!\ar^2 \ln\,\Ntil\, \nf \, \exp(2\ar\ca\ln^2 \Ntil)
\Big[\Big(4\:\!\ca-\:\!\cf\Big) + \frct{1}{3}\,\ar\ln^2 \Ntil\,
\ca\b0 \Big]\: . \quad
\label{eq:cLgNL}
\eea
where the first terms is the LL contribution. The resulting third-order 
$x$-space expression is
\bea
\label{cL3DL2}
 \x1^{-1}\: C_{L,{\rm g}}^{\,T}(x)\Big|_{\ar^{\,3}} & = & \:
     8\,\* \cf \* \cas\, \* \ln^{\,4}\! \x1 
\nn \\[-1mm] & & \mbox{\hspn}
     + \ln^{\,3}\! \x1\: \* \cf \* \Big [ 
                \frct{20}{3}\, \* \cfs 
          \,+\, \frct{52}{3}\, \* \cf \* \ca 
          \,-\, \frct{952}{9}\, \* \cas
          \,+\, \frct{16}{9}\, \* \ca \* \nf 
          \Big] \:\: + \:\: \dots \:\: . \quad
\eea
The third logs for $C_{L,{\rm g}}^{\,T}$ can not be derived by resumming the 
NLO results. The coefficient at order $\as^{\,3}$, however, can be obtained
be comparing the physical kernels for ($F_T,\,F_L^T$), cf.~Ref.~\cite{BRvN00}, 
to the analogous DIS results \cite{SMVVpr,ASV1} along the lines of 
Refs.~\cite{Grunberg:2011gx,AMV1}, yielding the continuation of 
Eq.~(\ref{cL3DL2})
$$
  \dots \;+\; \ln^{\,2}\! \x1\; \* \cf \* \Big[
             ( 62 - 32\,\*\z2\! )\, \* \cfs
           - \Big(\, \frct{784}{3}-32\,\*\z2\! \Big)\, \* \cf \* \ca
           + \frct{5720}{9}\,\* \cas
           + \frct{16}{3}\,\* \cf \* \nf 
           - \frct{224}{9}\,\* \ca \* \nf 
  \Big] \;+\; \dots 
$$
where, as in Ref.~\cite{SMVVpr}, an additional $\dabc2$ contribution has been 
suppressed for brevity. The derivation of this result and its extension to
the $\ln \x1$ term will be discussed in Ref.~\cite{ALPVprp}.
The numerical size of the resummed large-$N$ corrections is shown in Fig.~2,
again using $\Qs \simeq M_Z^{\,2}$.
  
\section{Discussion and Outlook}

\vspace*{-3mm}
\noindent
We have derived the all-order resummation of the three highest (NNL) threshold
double logarithms for the off-diagonal timelike splitting functions 
$P^{\,T}_{ij}$ and the coefficient functions $C_{T,g\,}^{\,T}$, 
$C_{L,g\,}^{\,T}$, $C_{\phi,\,q\,}^{\,T}$ and $C_{L,\,q\,}^{\,T}$ for 
gauge-boson and (in the heavy-top limit) Higgs exchange semi-inclusive $e^+e^-$ 
annihilation (SIA). Our results for the last quantity confirm the findings 
obtained in Ref.~\cite{MV5} by a different method, while the others are new. 
For brevity only a part of these results have been discussed here; for a full 
account the reader is referred to Ref.~\cite{ALPVprp}.

The numerical effect of the double logarithms beyond the third order on the 
splitting functions in Mellin-$N$ space is very small for a strong coupling 
$\as \simeq 0.12$ corresponding to a scale close to the $Z$-boson mass, 
amounting to less that 1\% for $N \lsim 20$. Here the contributions of terms 
beyond the fourth order are negligible.  
The corresponding corrections are larger, and receive noticeable contributions 
to order $\as^{\,5}$, for the coefficient functions. 

In (almost) all these cases these is no reason to believe that the N$^\ell$LL 
corrections for $\ell > 2$ are small compared to the present results. 
This is hardly surprising, given previous experience with other end-point 
resummations, but indicates that more terms are required in order to achieve 
phenomenological relevance. 
We hope that the present results will provide useful information for the 
development of more sophisticated approaches in the future.

As done at the fourth-order for non-singlet quark coefficient functions in 
Ref.~\cite{MV5} and the spacelike splitting functions in Ref.~\cite{SMVV1},
it is possible to extend the present results to all higher orders in the 
expansion in powers of $\x1$, i.e., to all terms of the form
$\x1^a \ln^{\,2n-\ell-n_0^{}}\x1$. Consequently all large-$x$ double logarithms in 
inclusive deep-inelastic scattering (DIS) and SIA
are fixed by lower-order information, with the coefficient of the N$^\ell$LL
contributions determined by the N$^\ell$LO fixed-order results.

Due to the somewhat different structure of the $D$-dimensional phase-space
integrations, the present approach is unfortunately not directly applicable 
(beyond the leading logarithms) to the Drell-Yan lepton-pair and Higgs-boson 
production in proton-proton collisions. Hence further, more refined tools are 
required in these cases to improve upon the physical-kernel constraints of 
Ref.~\cite{MV5} and to extend those results to the flavour-singlet 
contributions.

The present approach can be extended, on the other hand, to high-energy
(small-$x$) double logarithms, if neither to all quantities nor to all powers
$a$ in the analogous small-$x$ expansion in terms of 
$x^{\,a} \ln^{\,2n-\ell-n_0^{}} x$
terms. The NNLL resummation of the dominant $x^{\,-1}$ terms in the timelike
splitting functions and the fragmentation functions in SIA has been presented
in Ref.~\cite{AV2011}, the results for the $x^{\,0}$ terms for the 
corresponding DIS quantities have also been derived and will be presented
elsewhere \cite{KVprp}. Clearly more research is required on both the threshold
and high-energy~logarithms.

\vspace*{5mm}
\noindent
{\bf Acknowledgements}\\[1mm]
The participation of A. Lo Presti in this conference was made possible by
support from the Tata Institute of Fundamental Research, India, and the 
European-Union network {\it LHCPhenoNet} with contract number 
PITN-GA-2010-264564 of which also A. Almasy and A. Vogt are members.
The work of of A.A. has also been supported by the Deutsche 
Forschungsgemeinschaft in Sonderforschungsbereich/Transregio~9, and that of 
A.V. by the UK Science \& Technology Facilities Council (STFC) under grant 
number ST/G00062X/1.

\end{document}